\input{aipcheck}

\documentclass[
    ,final            
  ]
  {aipproc}

\layoutstyle{6x9}

\begin{document}

\title{Particle Acceleration at High-$\gamma$ Shock Waves}


\author{Jacek Niemiec}{
  address={Department of Physics and Astronomy, Iowa State University, Ames, 
  IA 50011, USA},
  altaddress={Institute of Nuclear Physics PAN, 31-342 Cracow, Poland}
}

\begin{abstract}
First-order Fermi acceleration processes at ultrarelativistic ($\gamma\sim$ 5-30)
shocks 
are studied with the method of Monte Carlo simulations. The accelerated particle 
spectra are obtained by integrating the exact particle trajectories in 
a turbulent magnetic field near the shock. The magnetic field model assumes 
finite-amplitude perturbations within a wide wavevector range and with a 
predefined wave power spectrum, which are imposed on the mean field component 
inclined at some angle to the shock normal. The downstream field structure is 
obtained as the compressed upstream field. We show that the main acceleration 
process at oblique shocks is the particle compression at the shock. Formation of 
energetic spectral tails is possible in a limited energy range for highly 
perturbed magnetic fields. Cut-offs in the spectra occur at low energies in 
the resonance range considered. We relate this feature to the structure of the 
magnetic field downstream of the shock, where field compression produces 
effectively 2D turbulence in which cross-field diffusion is very small. 
Because of the field compression downstream, the acceleration process is 
inefficient also in parallel high-$\gamma$ shocks for larger turbulence 
amplitudes, and features observed in oblique shocks are recovered.
For small-amplitude perturbations, particle spectra are formed in 
a wide energy range and modifications of the acceleration process due to the 
existence of long-wave perturbations are observed. The critical turbulence 
amplitude for efficient acceleration at parallel shocks decreases 
with $\gamma$.
We also study the influence of strong short-wave perturbations, generated 
downstream of the shock, on the particle acceleration processes at 
high-$\gamma$ shocks. The spectral indices obtained do not converge to the 
``universal'' value $\alpha\approx 4.2$. Our results indicate inefficiency of 
the first-order Fermi process to generate high-energy cosmic rays at 
ultrarelativistic shocks with the perturbed magnetic field structures 
considered in the present work.
\end{abstract}

\maketitle

\noindent
In the present work we investigate the first-order Fermi process
at relativistic shocks in test-particle approach with the method of Monte 
Carlo simulations. The applied magnetic field model includes the characteristic
features essential for a realistic description of the acceleration process.
The upstream magnetic field consists of the uniform component {\bf\em B}$_{0,1}$,
inclined at some angle $\psi_1$ to the shock normal, and static finite-amplitude 
perturbations imposed upon it. The irregular component has 
either a flat $(F(k)\sim k^{-1})$ or a Kolmogorov $(F(k)\sim k^{-5/3})$ wave 
power spectrum defined in a wide wavevector range with $k_{max}/k_{min}=10^5$.
The downstream field structure is obtained as the compressed upstream field,
so that the continuity of the turbulent magnetic field across the shock is
preserved. This allows one to study correlations in particle motion introduced
by the field structure for different levels of turbulence, and to investigate 
their influence on the particle spectra. Furthermore, the magnetic 
field model enables one to discuss the role of long-wave magnetic field
perturbations. The details of the model and earlier results for acceleration at
mildly relativistic shocks have been described elsewhere [1]. 
Here, we apply the model to investigate particle acceleration
at high-$\gamma$ shocks (see also [2]). The particle spectra are 
obtained by following exact particle trajectories in the perturbed magnetic field
near the shock, without the simplifying hybrid approach used in [1].
 
\begin{figure}
  \includegraphics[height=.29\textheight]{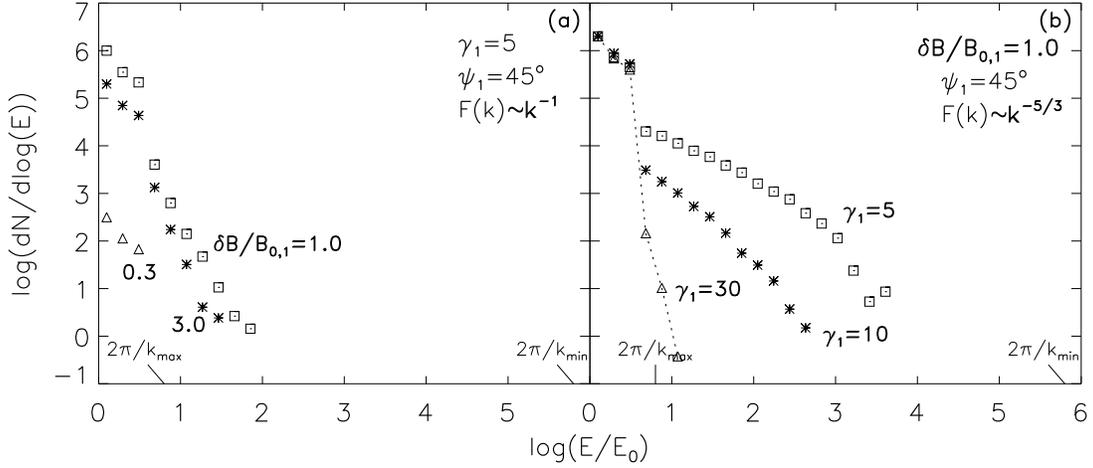}
  \caption{Accelerated particle spectra at oblique ($\psi_1=45^o$) superluminal 
shock waves in the shock normal rest frame. The shock is assumed to be a planar 
discontinuity, propagating with 
Lorentz factor $\gamma_1$ with respect to the upstream (electron-proton) plasma.
Figure 1a shows particle spectra for $\gamma_1=5$ and a flat power
spectrum of magnetic field perturbations for different upstream perturbation 
amplitudes $\delta B / B_{0,1}$. Spectra have vertical shifts for clarity.
Particle spectra obtained for different $\gamma_1$ and the Kolmogorov wave power 
spectrum with $\delta B / B_{0,1}=1.0$ are presented in Fig. 1b. 
Particles in the range ($2\pi/k_{max}$, $2\pi/k_{min}$) can satisfy the resonance
condition $k_{res}\simeq 2\pi /r_g(E)$.}
\end{figure}

The characteristic features of particle acceleration processes at oblique
superluminal high-$\gamma$ shocks are illustrated in Fig. 1. All injected 
particles are initially accelerated in a phase of ``superadiabatic'' compression 
at the shock [3]. Only a much smaller fraction of these particles is further 
accelerated in the first-order Fermi process, forming an energetic tail in the 
spectrum for highly perturbed magnetic fields. The shape of the spectral tail
and its extension to high particle energies strongly depend on the magnetic 
field turbulence spectrum. For the flat wave power spectrum, the tails are very
steep (Fig. 1a), whereas for the Kolmogorov turbulence, with most power in 
long-wave perturbations, they are much flatter and exhibit a continuous slow 
steepening (Fig. 1b). Cut-offs appearing in the spectra occur in the resonance 
energy range considered, and the cut-off energy decreases with growing shock 
Lorentz factor $\gamma_1$ (Fig. 1b).

The features observed in the spectra can be explained by the 
turbulent magnetic field structure at the shock. The field compression 
downstream of the shock produces effectively 2D turbulence in which particle 
diffusion conditions are highly anisotropic, with very limited diffusion along 
the shock normal. This leads to high particle escape rates and results in steep
particle spectra, as seen in Fig. 1a. Flatter particle spectra for the Kolmogorov 
turbulence (Fig. 1b) result from the effects of high-amplitude long-wave magnetic
field perturbations which can form locally subluminal field configurations at the
shock, thus enabling more efficient particle-shock interactions.
Increasing $\gamma_1$ leads to a decrease in the particle cross-field diffusion 
downstream of the shock, which is the most important factor leading to the 
lowering of the cut-off energy, as seen in Fig. 1b. 

The effects of downstream magnetic field compression may also occur in parallel 
high-$\gamma$ shocks. In the case of 
large-amplitude perturbations, the field compression leads to the effectively 
perpendicular shock configuration.  The acceleration processes thus become less 
efficient and features analogous to those observed in superluminal shocks are 
recovered. In conditions of weakly perturbed magnetic fields, wide-energy range 
particle spectra are formed. They are non-power-law in the entire energy range,
and their power-law parts are flat ($\alpha < 4$) due to the presence of 
long-wave perturbations, as reported previously for mildly relativistic 
shocks [1]. The critical turbulence amplitude for efficient acceleration 
at parallel shocks decreases with $\gamma_1$. This character of
the acceleration process deviates from the results of the classical models, e.g.,
[4-6], which suggest the existence of a ``universal'' spectral index 
$\alpha\approx 4.2$ for ultrarelativistic shocks.

\begin{figure}
  \includegraphics[height=.29\textheight]{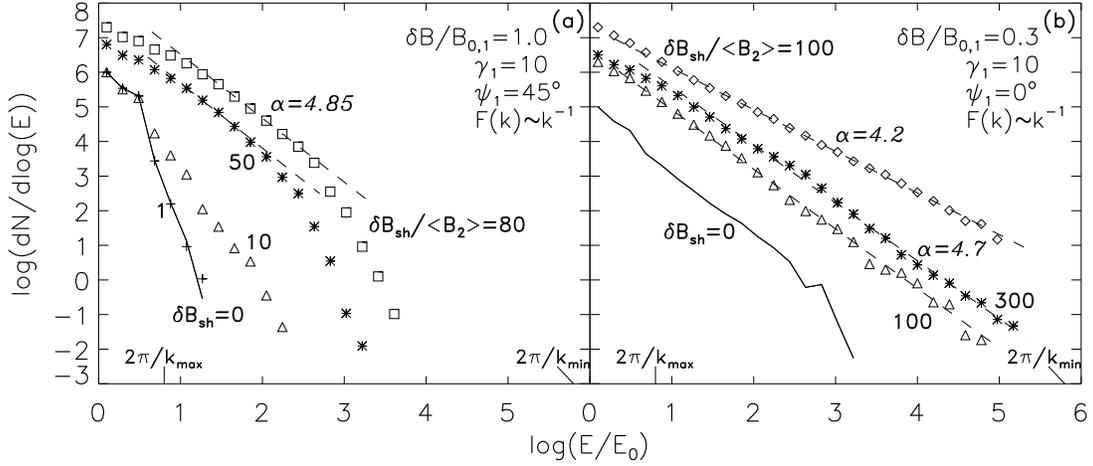}
  \caption{Accelerated particle spectra at ultrarelativistic shocks 
with $\gamma_1=10$ formed in 
the presence of small-scale large-amplitude perturbations generated downstream 
of the shock. The wave power spectrum of the compressed 
magnetic field component is flat ($F(k)\sim k^{-1}$) upstream of the shock.
Particle spectra at superluminal shocks ($\psi_1=45^o$) 
and $\delta B/B_{0,1}=1.0$ are shown in Fig. 2a, and spectra at parallel shocks
($\psi_1=0^o$) and $\delta B/B_{0,1}=0.3$ are presented in Fig. 2b.
Energy density in the short-wave component relative to the energy density in 
the compressed downstream magnetic field $\delta B_{sh}/<B_2>$ is given near 
the respective results. Linear fits to the spectra are presented and values
of the (phase-space) spectral indices $\alpha$ are given in italic 
(the energy spectral index $\sigma=\alpha -2$). The spectra shown with solid line
are obtained in the model without small-scale perturbations.
Spectrum shown with diamonds in Fig. 2b represents the model with particle 
pitch-angle scattering upstream of the shock, which does not include the effects 
of long-wave magnetic field perturbations.}
\end{figure}

We also study the effects of an additional highly nonlinear 
short-wave turbulence generated downstream of the shock due to, e.g., the Weibel
instability [7-9]. We assume that these short-wave
perturbations form isotropic 3D turbulence, which provides efficient particle 
scattering and may lead to a decorrelation between particle motion and the 
compressed field downstream of the shock. The influence of such perturbations on 
particle trajectories is included as a small-amplitude momentum scattering term, 
as in the hybrid method described in [1] (see also [10]) .

Particle spectra formed in the presence of the short-wave turbulence
downstream of the superluminal shock with $\gamma_1=10$ are shown in Fig. 2a. 
Efficient particle acceleration is possible if the energy density in the 
short-wave component is much higher than the mean energy density in the 
compressed downstream magnetic field. In such conditions, particle spectra with
(steep) power-law parts can be formed, and the spectral index does not depend on
$\delta B_{sh}/<B_2>$ above some critical value. Cut-offs in the spectra occur
because the efficiency of particle scattering due to small-scale perturbations 
diminishes with particle energy. This means that $\delta B_{sh}/<B_2>$
must be extremely large to decorrelate the motion of high-energy particles from 
the compressed downstream field and produce power-law spectra in 
the full energy range. 
Accelerated particle spectra formed at parallel shocks with $\gamma_1=10$ are 
shown in Fig. 2b. Note that the particle spectral index deviates from the 
``universal'' value $\alpha\approx 4.2$, even in the limit of 
$\delta B_{sh}/<B_2>\gg 1$. Only in the model with pitch-angle scattering 
upstream of the shock, which does not include the effects of long-wave 
perturbations, the ``universal'' spectral index is observed. 

Our results require a revision of many earlier discussions of cosmic 
ray acceleration up to very high energies in the first-order Fermi process 
at ultrarelativistic shocks. The modeling shows that for the analyzed turbulent 
magnetic field structures near the shocks, which are consistent with the shock
jump conditions, it is difficult to form wide-energy power-law particle spectra
in quasi-perpendicular (superluminal) and also parallel high-$\gamma$ shocks. 
This substantially modifies the shock acceleration picture of 
classical models which predict power-law spectra, often with the 
``universal'' spectral index. The presence of highly nonlinear short-wave
turbulence at the shock can lead to more efficient acceleration, but the energy
density in the small-scale component required to do so may be unrealistically 
high, in particular for large $\gamma_1$. The formation
of a spectrum with ``universal" index requires special conditions, which 
include strong particle scattering both downstream and upstream of the shock.
This suggests possibly different mechanism for particle acceleration to very high
energies, e.g., [11]. However, a detailed knowledge of the magnetic field 
structure near a relativistic shock is needed to reach definite conclusions as 
to the effectiveness of particle acceleration.


\noindent
The work was partially supported by the Polish State 
Committee for Scientific Research in 2002-2005 as a research project 
PBZ-KBN-054/P03/2001. 



\bibliographystyle{aipproc}   

\bibliography{sample}

\IfFileExists{\jobname.bbl}{}
 {\typeout{}
  \typeout{******************************************}
  \typeout{** Please run "bibtex \jobname" to optain}
  \typeout{** the bibliography and then re-run LaTeX}
  \typeout{** twice to fix the references!}
  \typeout{******************************************}
  \typeout{}
 }



\end{document}